\documentclass[twocolumn,preprintnumbers,amsmath,amssymbn,reprint,superscriptaddress]{revtex4}

\usepackage{graphicx}
\usepackage{dcolumn}
\usepackage{bm}
\usepackage{epstopdf}
\usepackage{float}
\usepackage{amsmath}

\usepackage[USenglish]{babel}
\addto{\captionsUSenglish}{}
\usepackage[utf8]{inputenc} 
\usepackage[hidelinks]{hyperref}
\usepackage{amssymb,amsmath}

\newcommand{\ket}[1]{| #1\rangle}

\newcommand*\colvec[3][]{\begin{pmatrix}\ifx\relax#1\relax\else#1\\\fi#2\\#3\end{pmatrix}}

\newcommand{\kk}{\mathbf{k}}
\newcommand{\qq}{\mathbf{q}}

\newcommand*\dd{\mathop{}\!\mathrm{d}}

\begin{document}
\title{Interference Fringes Controlled by Non-Interfering Photons}

\author{Armin Hochrainer}
\email{armin.hochrainer@univie.ac.at}
\affiliation{Institute for Quantum Optics and Quantum Information, Austrian Academy of Sciences, Boltzmanngasse 3, Vienna A-1090, Austria.}
\affiliation{Vienna Center for Quantum Science and Technology (VCQ), Faculty of Physics, Boltzmanngasse 5, University of Vienna, Vienna A-1090, Austria.}

\author{Mayukh Lahiri}%
\affiliation{Institute for Quantum Optics and Quantum Information, Austrian Academy of Sciences, Boltzmanngasse 3, Vienna A-1090, Austria.}
\affiliation{Vienna Center for Quantum Science and Technology (VCQ), Faculty of Physics, Boltzmanngasse 5, University of Vienna, Vienna A-1090, Austria.}

\author{Radek Lapkiewicz}
\affiliation{Institute for Quantum Optics and Quantum Information, Austrian Academy of Sciences, Boltzmanngasse 3, Vienna A-1090, Austria.}
\affiliation{Vienna Center for Quantum Science and Technology (VCQ), Faculty of Physics, Boltzmanngasse 5, University of Vienna, Vienna A-1090, Austria.}
\author{Gabriela B. Lemos}
\affiliation{Institute for Quantum Optics and Quantum Information, Austrian Academy of Sciences, Boltzmanngasse 3, Vienna A-1090, Austria.}
\affiliation{Vienna Center for Quantum Science and Technology (VCQ), Faculty of Physics, Boltzmanngasse 5, University of Vienna, Vienna A-1090, Austria.}
\author{Anton Zeilinger}
\email{anton.zeilinger@univie.ac.at}
\affiliation{Institute for Quantum Optics and Quantum Information, Austrian Academy of Sciences, Boltzmanngasse 3, Vienna A-1090, Austria.}
\affiliation{Vienna Center for Quantum Science and Technology (VCQ), Faculty of Physics, Boltzmanngasse 5, University of Vienna, Vienna A-1090, Austria.}

\date{\today}

\begin{abstract}

We observe spatial fringes in the interference of two beams, which are controlled by a third beam through the phenomenon of induced coherence without induced emission.
We show that the interference pattern depends on the alignment of this beam in an analogous way as fringes created in a traditional division-of-amplitude interferometer depend on the relative alignment of the two interfering beams. We demonstrate that the pattern is characterized by an equivalent wavelength, which corresponds to a combination of the wavelengths of the involved physical light beams.

\end{abstract}

\maketitle

The relationship between path-information and interference is fundamental to quantum physics \cite{feynman_quantum_1965} and has been studied in various contexts \cite{greenberger_simultaneous_1988,mandel_coherence_1991,englert_fringe_1996}. A particularly remarkable manifestation of this relationship is the phenomenon of induced coherence without induced emission \cite{wang_induced_1991,zou_induced_1991}. This phenomenon was used for fundamental tests of complementarity \cite{herzog_frustrated_1994,herzog_complementarity_1995,heuer_complementarity_2015} and led to applications in imaging \cite{lemos_quantum_2014,lahiri_theory_2015}, metrology \cite{hudelist2014quantum}, spectrum shaping \cite{iskhakov2016nonlinear}, and spectroscopy \cite{kulik2004two,kalashnikov_infrared_2016}.
If two spatially separated nonlinear crystals produce photon pairs (signal and idler) by spontaneous parametric down-conversion (SPDC) \cite{klyshko_scattering_1969,burnham_observation_1970}, the emitted signal beams in general do not interfere, even if the two pump beams are mutually coherent.
This is due to the fact that the idler beams carry information which source a down-converted pair originated from.
By aligning the idler beams emitted by the two crystals,
this information can be suppressed. If the path lengths are chosen accordingly, then lowest-order interference is observed between the two signal beams. 
It has been shown that path distinguishability can be introduced through a time delay between the two idler beams \cite{zou_control_1993} or by attenuating the idler beam from one source using a partially transmissive filter.
In the latter case, the resulting visibility was quantitatively connected to the transmission coefficient \cite{zou_induced_1991,wang_induced_1991}.

Here, we analyze a situation, in which distinguishability is introduced in a different way. This is done by marginally misaligning the idler beams, either by tilting or by defocusing one with respect to the other.
One could conjecture that this has the analogous effect of merely a reduced visibility.
However,
our experiment shows that because the misalignment can be described by a transverse phase gradient, it results in the observation of spatial interference fringes. This interference pattern leads to a reduction of visibility when intensities are obtained by integrating over a transverse section of the beam.

We demonstrate analogies and differences between these fringes created by \emph{induced} coherence 
and those created in traditional interferometry by division of amplitude, i.e. by interfering two beams, which are derived from a single beam by splitting it on a semi-reflecting surface.
In a standard division-of-amplitude interferometer (e.g. a Michelson interferometer), a fringe pattern is observed, when one of the two interfering beams is misaligned with respect to the other. In particular, parallel or round fringes occur, depending on whether the tilt or the propagation distance of one beam is changed.
In our case, the fringe pattern exhibits an analogous structure if we keep the two interfering signal beams aligned, but misalign the idler beam from one crystal with respect to the other idler beam.

We further show that the obtained fringe pattern is characterized neither by the signal nor by the idler wavelength alone, but by a combination of the involved wavelengths. We attribute this effect to the momentum correlation in non-degenerate SPDC.

\begin{figure}[t]
\includegraphics[width=\linewidth]{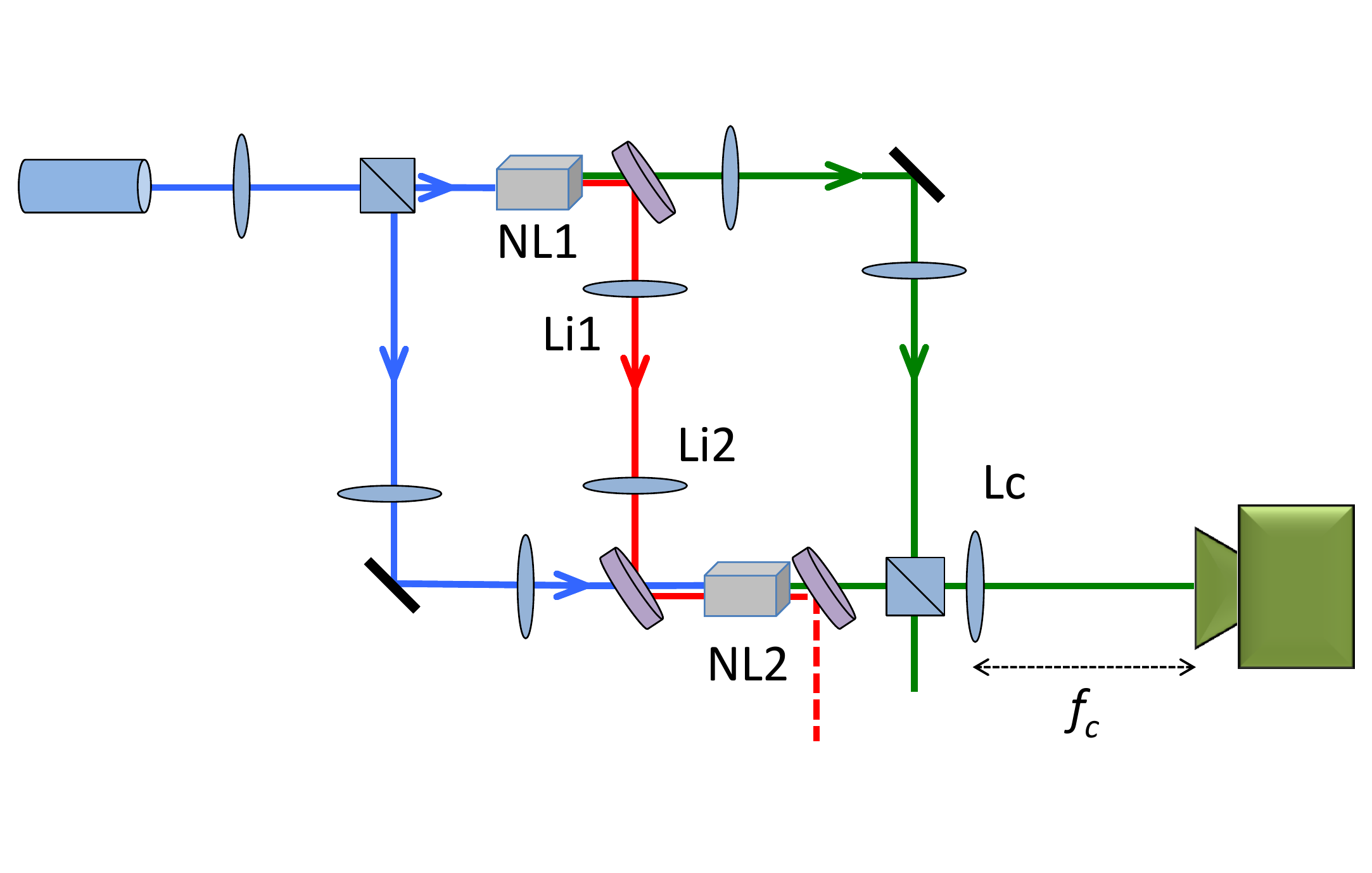}
\caption{Setup of the experiment. A continuous-wave pump laser (blue, 532 nm) is split at a beam splitter and focused into two nonlinear crystals NL1 and NL2. The crystals produce signal (green, 810 nm) and idler (red, 1550 nm) photon pairs by type-0 SPDC. The idler beam from NL1 is overlapped with the idler beam from NL2 such that the two beams are indistinguishable after NL2. The two signal beams are superposed at a beam splitter and a camera detects the output at the focal distance of a positive lens (Lc). Confocal lens systems ensure identical pump spots at the crystals and indistinguishability of each mode of the idler beams at NL2. Coherence is induced between the respective modes of the signal beams. The lens systems further assure that the effect of the longer path-length of the signal beam from NL1 is canceled.
}
\label{fig:setup}
\end{figure}

Our setup is depicted in Fig. \ref{fig:setup}. Two nonlinear crystals (NL1 and NL2) are pumped by the same laser source and produce photon pairs by non-degenerate SPDC. The two signal beams are superposed at a beam splitter and subsequently detected by a camera.
Lens systems assure that if the signal fields at the two crystals are decomposed into plane wave components, one wave-vector (spatial mode) of each signal field is detected at one point on the camera. In other words, the camera detects the transverse Fourier transform of the superposed signal field.
The idler beam emerging from NL1 is directed through NL2 and aligned with the idler beam generated at NL2. A 4$f$ lens system is inserted between the two crystals, which produces an image of the idler beam cross-section from NL1 at NL2.
In this way, each spatial mode of the idler beam after NL2 has almost equal probability of being populated with an idler photon generated at NL1 or with an idler photon from NL2.
If the path lengths are chosen accordingly, an idler photon detected after NL2 does not carry any information about which crystal it emerged from. Therefore, it is impossible to infer from which crystal its partner signal photon arrives at the beam splitter. As a consequence, interference between the two signal beams can be observed in each spatial mode, i.e. at each point on the camera. No heralding or coincidence detection is used.
If the idler beam is blocked between NL1 and NL2, the signal beams do not interfere.
The pair production rate in our experiment is low enough to render the possibility of stimulated emission negligible, as the probability of two idler photons being present in the setup at the same time is practically zero.

\begin{figure}[b]
\centering
\includegraphics[width=.45\textwidth]{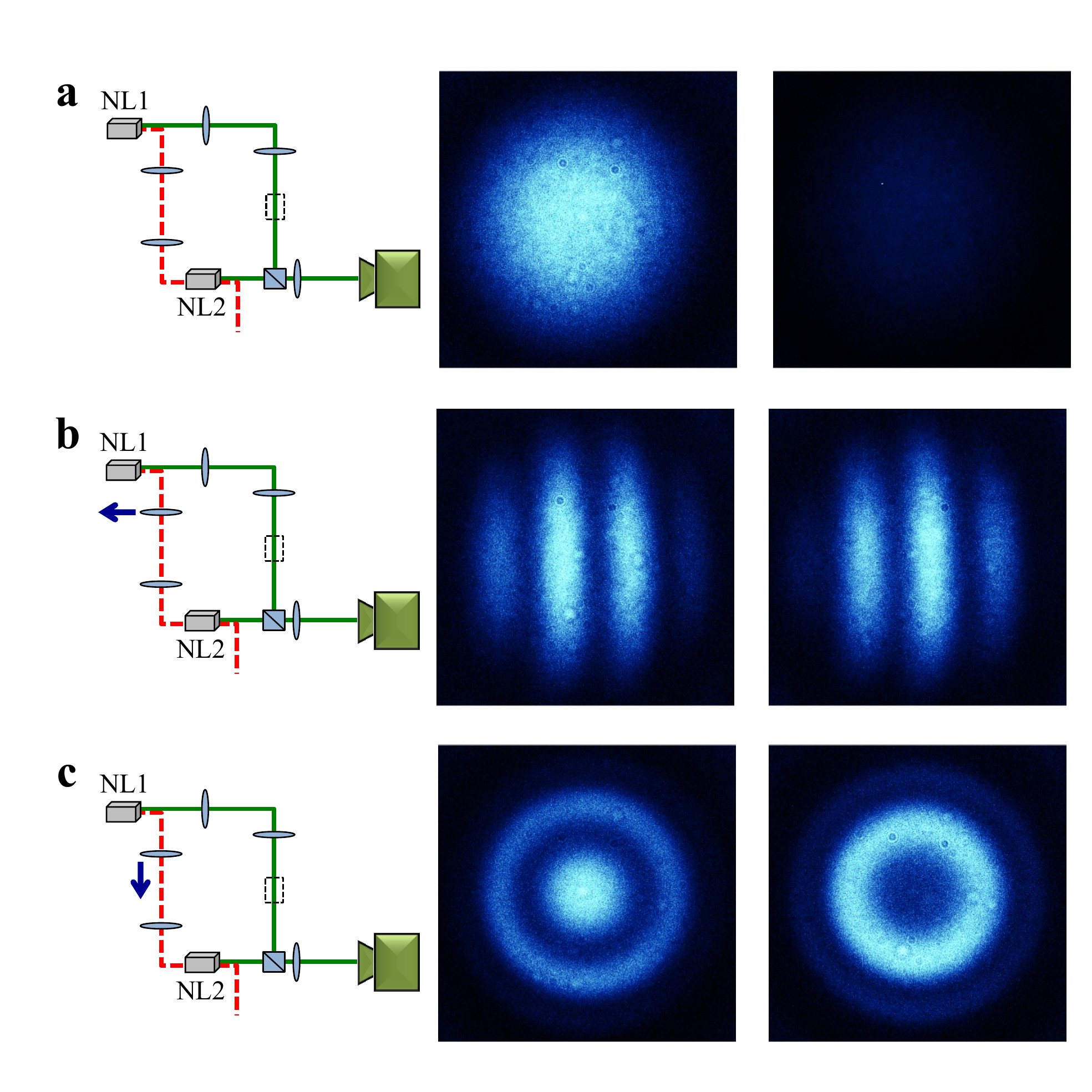}
\caption{Interference fringes in the signal beam for different idler beam configurations. (a) On perfect alignment of the 4$f$ imaging system on the idler beam, a uniform interference pattern is observed as a bright or a dark spot on the camera as the phase is shifted by $\pi$. (b) By tilting the idler beam, parallel fringes are produced on the camera. (c) If the imaging system is defocused, circular fringes occur.}
\label{fig:3scenarios-neuwebilder-bitmap}
\end{figure}

The lens systems effectively cancel the effect of free-space propagation between the crystals. In the case of perfect alignment, the observed interference pattern exhibits a uniform intensity modulation along the beam cross section. That is, by scanning the position-independent interferometric phase, all points on the camera simultaneously undergo the same transition from maximum to minimum intensity (See Fig. \ref{fig:3scenarios-neuwebilder-bitmap}a). Note that the interferometric phase can be modulated by changing the optical path length of either one of the signal beams, the idler beam between NL1 and NL2, or by changing the relative phase of the pump beam between NL1 and NL2. In our experiment, we investigate situations, in which the relative alignment of the two idler beams is modified. Maintaining perfect alignment of the two signal beams, a small tilt of the idler beam generated at NL1 relative to that generated at NL2 causes parallel interference fringes to appear (Fig. \ref{fig:3scenarios-neuwebilder-bitmap}b). This tilt is implemented by translating a lens of the imaging system perpendicular to the beam propagation axis. A different interference pattern is created if instead of transversely displacing the idler beam, the imaging system from NL1 to NL2 is defocused.

The observation of a spatially dependent interference pattern in the superposition of signal beams can be understood considering the quantum state of light in our experiment.
Each crystal NL$j$ produces photon pairs in a set of spatial modes labeled by the wave vectors $\kk_S$ and $\kk_I$,
\begin{equation}
\ket{\psi_{\text{NL}j}}=\int \dd\kk_{S_j} \dd\kk_{I_j} C(\kk_S,\kk_I)\ket{\kk_S}_j\ket{\kk_I}_j .
\end{equation}
The coefficients $C(\kk_S,\kk_I)$ determine the properties of the emission. In our experiment, two crystals are pumped coherently and emit into identical idler modes. 
This is represented by re-labeling idler modes from different crystals in the superposition $\ket{\psi_{\text{NL}1}}+\ket{\psi_{\text{NL}2}}$ as ${\kk_I}_1\rightarrow{\kk_I},{\kk_I}_2\rightarrow{\kk_I}$ (a more detailed theoretical treatment can be found in \cite{fringetheory}). The projection of the final bi-photon state onto one particular mode $\kk_S'$ of the signal beam is given by
\begin{align}
\ket{\Psi_{\kk_S'}}=\int & \dd\kk_I C(\kk_S',\kk_I)& \nonumber \\
&\times \bigg(\ket{\kk_S'}_1+ e^{i(\phi_I(\kk_I)+\phi_0)}\ket{\kk_S'}_2\bigg)\ket{\kk_I},
\label{state}
\end{align}
where $\phi_I(\kk_I)$ denotes the phase shift introduced in the idler mode $\kk_I$ between NL1 and NL2 and all other phase terms are combined in $\phi_0$. Since the lens in front of the camera maps one wave vector to one point on the camera, this state determines the detection probability at the corresponding point on the camera.

We consider the beams to be paraxial and the detection plane perpendicular to the optical axis. The superposed signal beam is detected after a 3 nm bandpass filter centered at the signal wavelength $\lambda_S$. We assume a well-collimated pump beam \footnote{In our experiment, the Gaussian waist of the pump beam was approximately 250 $\mu$m at a crystal length of 2 mm.} and neglect any transverse phase mismatch. In this case, the transverse wave vectors $\qq_S$ and $\qq_I$ of a signal and an idler photon belonging to the same pair are related by the phase matching condition
\begin{equation}
\qq_S+\qq_I=\qq_P\approx 0,
\label{phasematching}
\end{equation}
where $\qq_P$ denotes a transverse wave vector of the pump field.
The transverse momentum correlation implied by Eq. (\ref{phasematching}) is reflected in the coefficients $C(\kk_S,\kk_I)$. 
It follows from  Eq. (\ref{state}) that in this case, a spatially dependent phase introduced in the idler beam between the two crystals is observed in the interference pattern of the two signal beams.
Note that such spatial fringes can occur only if some correlation between the momenta of signal and idler photons exists \footnote{
In the case of uncorrelated momenta of signal and idler photons, $C(\kk_S,\kk_I)$ can be separated as $C_S(\kk_S)C_I(\kk_I)$. In this case, all $\kk_I$ contributing to the phase term in Eq. (\ref{state}) are integrated over, which diminishes the visibility of the spatial fringes.
}. The relationship between momentum correlation and fringe visibility is subject of a separate paper \cite{momcorr}.

\begin{figure*}[t]
\centering
\includegraphics[width=0.95\textwidth]{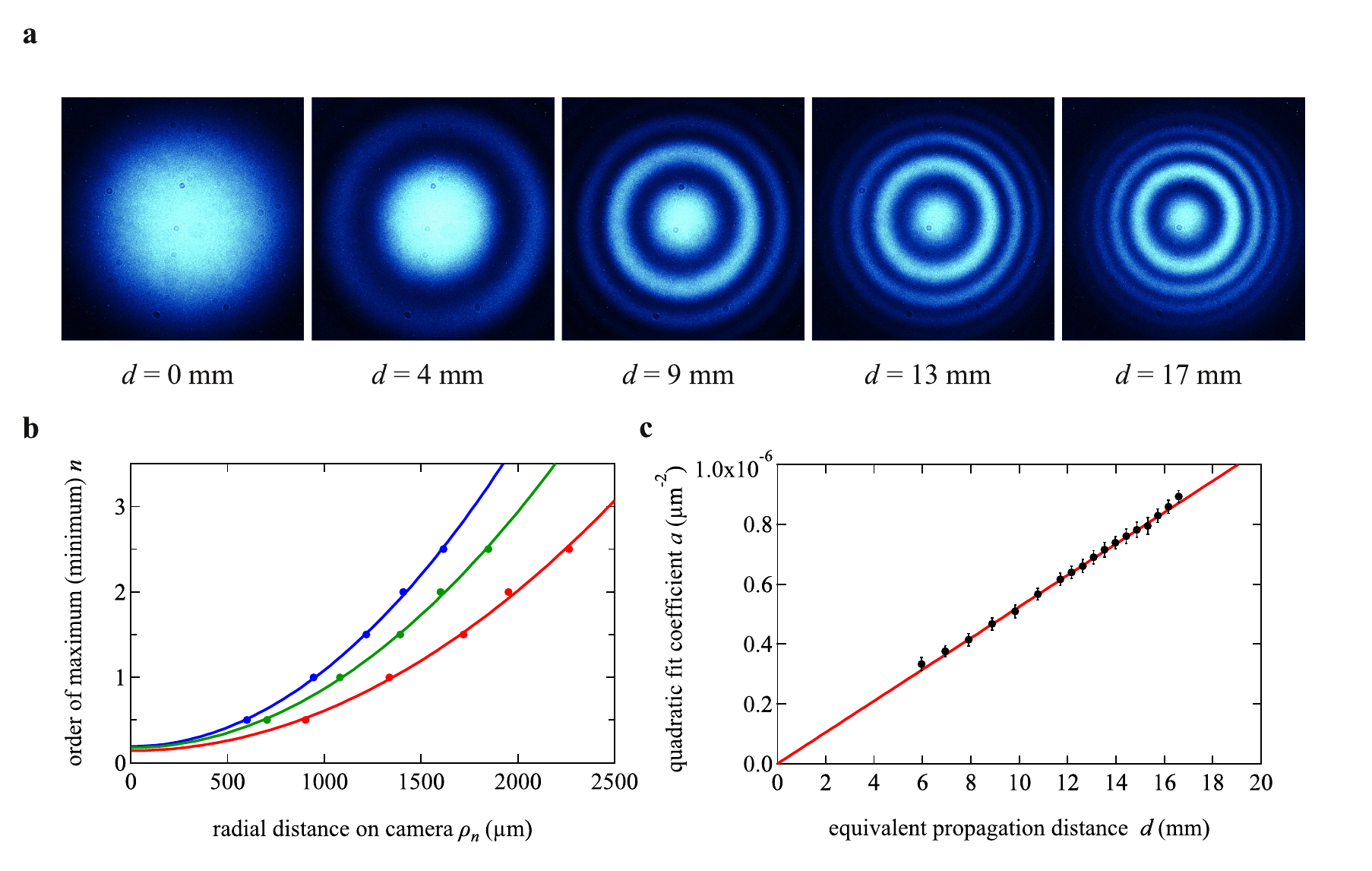}
\caption{Wavelength dependence of the interference pattern 
(a) Examples of camera images obtained by introducing effective propagation distances $d$ in the idler beam between the two crystals. (b) Evaluated radial positions $\rho_n$ of minima and maxima for $d$ = 9 mm (red), $d$ = 13 mm (green),and $d$ = 17 mm (blue). Error bars representing the standard deviation of the mean are smaller than the dots.
These radii are subject to the condition $a \rho_n^2+\phi_0= n$, where
$n$ represents the order of maxima ($n$ = 1, 2) and minima ($n$ = 0.5, 1.5, 2.5). The solid lines are parabolic fits to the data, used to determine $a$. (c) The quadratic coefficients $a$ (black points) determine how fast the relative phase between the two interfering beams varies with the distance from the center of the pattern. They show the expected linear dependence on the introduced propagation distance. The slope corresponds to an equivalent wavelength of $420 \pm 7$ nm. The solid line shows the theoretical prediction for an equivalent wavelength of $423$ nm.}
\label{fig:equivalentWL1}
\end{figure*}

In common optical interferometers, in which two interfering beams are created by division of amplitude \cite[Ch.7]{born_principles_1999}, circular fringes occur if the optical path length of one of the beams is extended with respect to the other. These fringes are often referred to as fringes of equal inclination or Haidinger fringes \cite[Ch.2]{hariharan_optical_2003}. Their transverse spacing at a given relative distance is determined by the wavelength of the interfering light.
In our interferometer, the wavelengths of signal ($\lambda_S =810$ nm) and idler ($\lambda_I=1550$ nm) beams differ significantly from each other. This gives rise to the question, which wavelength governs the spatial fringes in our experiment.

In order to create the fringes, the imaging system in the idler beam was defocused by translating a lens (Li1 in Fig. \ref{fig:setup}) about a distance $\delta$ along the beam propagation axis. The resulting phase shift can be approximated by the phase shift introduced by free-space propagation about a distance  $d=(f^2\delta)/(f^2+\delta^2)$ (supplementary). Here, $f$ denotes the focal length of the lenses in the idler beam (Li1 and Li2 in Fig. \ref{fig:setup}). Under this approximation, the introduced phase shift can be expressed as
\begin{equation}
\phi(\theta_I)=\frac{2 \pi}{\lambda_I}d\frac{\theta_I^2}{2},
\label{phaseshift}
\end{equation}
where $\theta_I$ is the angle, a wave vector of the idler beam subtends with the optical axis.
Due to the symmetry of Eq. (\ref{phaseshift}) in rotations about the optical axis, the fringes introduced in this way are circular.

As our sources produce non-degenerate photon pairs, correlated signal and idler plane wave components leave the crystals at different angles from the optical axis ($\theta_S$ and $\theta_I$ respectively), as a consequence of the phase matching condition, Eq. (\ref{phasematching}). For small angles, $\theta_S$ and $\theta_I$ are related to the wavelengths of signal and idler beams as 
\footnote{
This relation is obtained from Eq. (\ref{phasematching}) considering the refraction at the crystal surface \cite{lahiri_theory_2015}.
}
\begin{equation}
\frac{\theta_S}{\theta_I}\approx \frac{\lambda_S}{\lambda_I}.
\label{angles}
\end{equation}

Wave vectors $\theta_S$ of the superposed signal beam are observed on the camera at a transverse distance $\rho \approx f_c\theta_S$ from the optical axis. Here $f_c$ denotes the focal length of the lens in front of the camera (Lc in Fig. \ref{fig:setup}).
As a consequence of Eq. (\ref{state}) with the phase-matching condition Eq. (\ref{phasematching}), the phase introduced in the idler beam at $\theta_I$ modulates the intensity in the superposed signal beam at $\theta_S$. Equation (\ref{angles}) shows that this modulation is observed in the camera at the transverse distance from the beam center $\rho \approx f_c\theta_I\lambda_S/\lambda_I$.
Therefore, the observed interference pattern depends on both signal and idler wavelengths.

It can be shown that the radii of the resulting circular fringes on the camera obey the following condition for maxima and minima (cf. \cite{fringetheory}),
\begin{equation}
\frac{d}{2f_c^2}\rho_{n}^2+\varphi=n\frac{\lambda_S^2}{\lambda_I},
\label{minmaxcondition}
\end{equation}
where intensity maxima correspond to integer $n=0,1,2,3,$… and minima to $n = 0.5,1.5,2.5$…. Here, $\varphi$ is a phase-offset, which is constant across the beam cross-section. Equation (\ref{minmaxcondition}) closely resembles the condition for maxima and minima of fringes of equal inclination in classical interferometry \cite[Ch.7]{born_principles_1999}. However, the fringes in our experiment are characterized by an ``equivalent wavelength",
\begin{equation}
\lambda_{eq}=\frac{\lambda_S^2}{\lambda_I},
\end{equation}
which governs the number of minima and maxima within a certain radial distance on the camera for a given value of $d$.

Figure \ref{fig:equivalentWL1}a shows examples of resulting fringe patterns at the camera when phase-shifts corresponding to different propagation distances $d$ are introduced. The images were analyzed using a computer algorithm, which evaluated the radii $\rho_n$ of bright and dark fringes (circles of maximum and minimum intensity).
It follows from Eq. (\ref{minmaxcondition}) that these radii are subject to the condition $a \rho_n^2+\varphi'= n$, where $a=d/(2 f_c^2\lambda_{eq})$.
Figure \ref{fig:equivalentWL1}b shows experimentally obtained pairs of $(\rho_n,n)$ for three different values of $d$.
The coefficient $a$ was evaluated using 2nd order polynomial fits to the data.
It can be interpreted as a measure of the spatially dependent change of the relative phase between the two interfering beams.
In Fig. \ref{fig:equivalentWL1}c, the obtained values of $a$ are plotted for different propagation distances $d$ in comparison to the theoretical prediction. The equivalent wavelength was determined from the dependence of $a$ on $d$. The result, $\lambda_{eq} = 420 \pm 7$ nm agrees well with the theoretical value of 423 nm. 

This result shows that the circular fringe pattern created through induced coherence without induced emission is governed by a combination of signal and idler wavelengths and not by either of the two alone. This is a signature of the momentum correlation between signal and idler beams.

By measuring the equivalent wavelength, it is possible to obtain quantitative information about the idler photons from the spatial structure of the fringes in the signal beam. In particular, if the signal wavelength is known, we can determine the wavelengths of both idler and pump beams from the fringe pattern. No idler photon needs to be detected for this measurement.
For our choice of pump wavelength and crystal parameters, the equivalent wavelength is smaller than any of the involved physical wavelengths (of signal, idler or pump). This makes it potentially interesting for applications in metrology.

The role of different propagation distances of the two idler beams was studied previously in a setup without lens systems \cite{grayson_spatial_1994}. A reduction of visibility was predicted as a consequence of the distinguishability arising from a difference between the divergences and transverse sizes of the two aligned idler beams.
We have shown that the introduction of a propagation phase in the idler beam between the two crystals leads to spatial interference fringes. It is clear that if intensity measurements are performed by integrating over a finite cross-section of the beam, these spatial fringes lead to a reduced visibility. 

The appearance of circular fringes governed by the equivalent wavelength is not limited to the quantum effect of induced coherence without induced emission.
One could envisage a similar experiment in which the emission of signal photons is stimulated by a common light source. This could be done for example by sending a laser at the idler wavelength through the common idler path (see also \cite{wang_observation_1991}).
In this case, the two emitted signal beams would display many of the observed features.
In particular, similar fringes could be observed if an appropriate phase shift is introduced in the stimulating laser beam between the two crystals.
However, if the stimulating laser beam is blocked between the two crystals, no stimulated emission can occur in the second crystal, leading to a reduced emission rate of signal photons there. This is an important difference to our experiment, where the intensities of the individual signal beams do not change, when the idler beam is either blocked or misaligned.

The formation of striped and circular interference fringes is familiar from traditional interferometry, where similar patterns are produced e.g. by introducing a tilt or an additional propagation distance in one of the interfering beams. In this case, the fringe pattern is characterized by the wavelength of the interfering beams.
In our experiment, the interference pattern is produced by manipulating only the undetected idler beam. None of the two interfering beams traverses the lens system, which is used to produce and control the fringes. Nevertheless, the obtained fringe pattern resembles that of a traditional interferometer, in which the same manipulation is performed in one of the interfering beams. However, in contrast to a traditional interferometer, the pattern is characterized by a combination of the wavelengths of both photons.

As long as the idler beam in our experiment is blocked or not aligned, the signal beams are mutually incoherent and do not interfere. Therefore, it is impossible to attribute a deterministic phase difference to the two signal beams.
However, the alignment of the respective idler beams induces coherence between the signal beams.
The spatial dependence of the induced phase difference can be controlled in the undetected idler beam in an analogous way as a spatially dependent phase-shift can be controlled in the interfering beams of a classical interferometer.

\section*{Acknowledgements}
The authors thank F. Steinlechner for helpful discussions. This work was supported by the Austrian Academy of Sciences (\"OAW) - IQOQI Vienna and the Austrian Science Fund (FWF) with SFB F40 (FOQUS) and W1210-2 (CoQuS).

\let\Section\section 
\def\section*#1{\Section{#1}} 
\bibliographystyle{unsrt}

\clearpage

\subsection{SUPPLEMENTARY INFORMATION}

\begin{figure}[htbp]
\centering
\includegraphics[width=.45\textwidth]{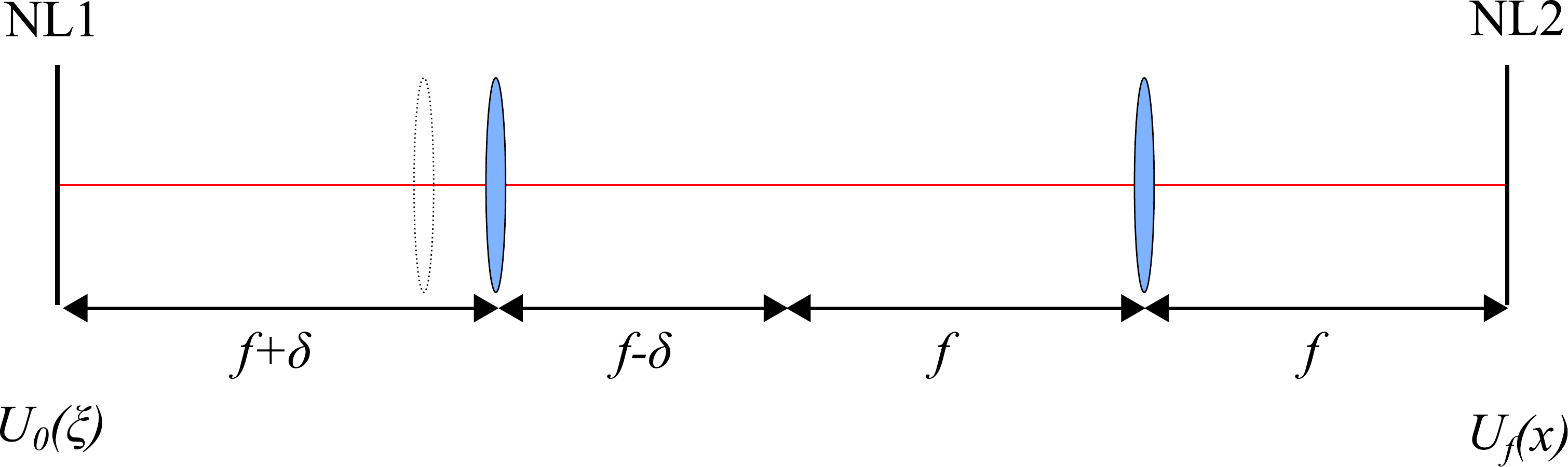}
\caption{Displaced 4$f$ lens system in the idler beam. See text.}
\label{fig:idlerlenssystem}
\end{figure}

Here, we justify why the translation of a lens in the 4$f$ imaging system in our experiment produces a phase shift equivalent to that introduced by free-space propagation.
The lens system through which the idler beam propagates between the two crystals is sketched in Fig. \ref{fig:idlerlenssystem}. The displacement of the first lens from its initial position in the 4$f$ imaging system is denoted by $\delta$. Due to the symmetry of our experimental setup, we restrict ourselves to a one-dimensional calculation.
The optical field of the idler beam at the plane in the center of NL1 is denoted by $U_0(\xi)$, where $\xi$ is the transverse position coordinate.
The optical field at the plane in the center of NL2 ($U_f(x)$) is obtained by Fourier optics calculation as

\begin{align}
U_f(x)\propto\int\limits_{-\infty}^{\infty} d\xi~ U_0(\xi)\exp\big(\frac{ik}{2}&\bigg[x^2\bigg(\frac{1}{\delta} 
+\frac{\delta}{f^2}\bigg) \nonumber \\
&-2x\xi\frac{1}{\delta}+\xi^2\frac{1}{\delta}\bigg]\big).
\label{idlerfielddisplacedlens}
\end{align}

Consider now a free-space propagation of the field $U_0(\xi)$ by a distance $d=f^2\delta/(f^2+\delta^2)$. In this case, the propagated field is given by \cite{goodman_introduction_2005}
\begin{align}\label{idlerfieldpropagation}
U_{prop}(x)\propto\int\limits_{-\infty}^{\infty}&d\xi~  U_0(\xi) \exp\big(\frac{ik}{2}\bigg[x^2\left(\frac{1}{\delta}+\frac{\delta}{f^2}\right) \nonumber \\
&-2x\xi\left(\frac{1}{\delta}+\frac{\delta}{f^2}\right)+\xi^2\left(\frac{1}{\delta}+\frac{\delta}{f^2}\right)\bigg]\big)
\end{align}
Comparing the terms in Eq. (\ref{idlerfielddisplacedlens}) and Eq. (\ref{idlerfieldpropagation}), it is clear that the two expressions are approximately equal when $\delta^2/f^2 << 1$. In our experiment, the largest value of $\delta^2/f^2$ is below 0.06.

\end{document}